\documentclass{article}

\usepackage{arxiv}

\usepackage[utf8]{inputenc} % allow utf-8 input
\usepackage[T1]{fontenc}    % use 8-bit T1 fonts
\usepackage{hyperref}       % hyperlinks
\usepackage{url}            % simple URL typesetting
\usepackage{booktabs}       % professional-quality tables
\usepackage{amsfonts}       % blackboard math symbols
\usepackage{nicefrac}       % compact symbols for 1/2, etc.
\usepackage{microtype}      % microtypography
\usepackage{lipsum}		% Can be removed after putting your text content
\usepackage{graphicx}
\usepackage[numbers]{natbib}
\usepackage{doi}

\title{KMS Structure, Detector Ordering, and Modular Irreversibility for Accelerated Quantum Systems}

\author{
Marcello Rotondo\thanks{Independent researcher. Contact: \texttt{marcello@gravity.phys.nagoya-u.ac.jp}}
}

\begin{document}
\maketitle

\begin{abstract}

We investigate when temporal ordering becomes physically meaningful in relativistic quantum field theory using localized detector models. A time parameter alone does not guarantee that different sequences of operations correspond to distinguishable processes. We show that distinguishability arises when two conditions are satisfied: the quantum state exhibits Kubo--Martin--Schwinger (KMS) structure and the detector couples through non-commuting observables.

For uniformly accelerated detectors in the Minkowski vacuum, the restriction of the field state induces a thermal response characterized by the Unruh temperature and the Tolman redshift profile. Sequential couplings through distinct observables produce an ordering-dependent contribution to the reduced detector state already at second order in perturbation theory. This contribution is governed by the symmetrized correlation function and acquires a universal thermal weighting fixed by the KMS condition.

The resulting asymmetry is quantified using relative entropy between effective Gibbs states associated with non-commuting generators. In a minimal two-level model, this yields a closed-form expression depending only on the dimensionless parameter set by the local temperature and detector energy scale. The comparison between Bogoliubov--Kubo--Mori and Bures metrics distinguishes irreversible entropic cost from operational distinguishability.

These results provide an operational realization of thermal time as modular flow normalized by the local temperature. The direction of thermal time is identified with the orientation for which relative entropy production is non-negative, linking KMS thermality, modular structure, and the emergence of a thermodynamic arrow in accelerated quantum systems.

\end{abstract}


\section{Introduction}

Uniformly accelerated observers provide one of the clearest settings in which the relation between quantum field theory, thermality, and spacetime structure becomes operational. The Unruh effect shows that the Minkowski vacuum, when probed along a uniformly accelerated trajectory, gives rise to a thermal detector response with temperature proportional to the proper acceleration \citep{Unruh1976,Davies1975,Crispino2008}. This result is not merely a statement about detector excitation probabilities. It reflects the Kubo--Martin--Schwinger (KMS) structure of vacuum correlation functions restricted to accelerated motion, and is closely related to the modular properties of wedge algebras in relativistic quantum field theory \citep{Haag1996,BratteliRobinson1997,Bisognano1975,Bisognano1976}.

The present paper investigates a question that arises naturally from this perspective. Given a detector moving along an accelerated trajectory, when does the time ordering of localized interactions become physically encoded in the detector state? The existence of a proper-time parameter in the interaction Hamiltonian is not sufficient by itself. If different interaction orderings lead to the same reduced state, the parameter orders the calculation but not an observable distinction. We therefore ask under what conditions the ordering defined by the detector proper time becomes operationally meaningful.

Detector models provide a direct way to address this question. In the standard Unruh--DeWitt model, a localized quantum system couples to a quantum field along a prescribed worldline, and its response is determined by the pullback of the field two-point function to that trajectory \citep{DeWitt1979,BirrellDavies1982,Takagi1986,LoukoSatz2006}. Standard analyses usually focus on a single detector monopole operator and on transition probabilities or response rates. Here we instead focus on the internal algebra of the detector and allow the detector to couple sequentially through distinct non-commuting observables.

This modification isolates a simple mechanism. Non-commuting detector observables make the order of operations potentially observable, while the KMS condition supplies a thermal relation between positive and negative frequency components of the field correlations. When both structures are present, reversing the order of the detector couplings changes the reduced detector state. The resulting difference is controlled by the same inverse temperature that appears in the Unruh effect.

The second part of the paper connects this ordering asymmetry with the modular structure of effective detector Gibbs states. For a two-level detector with Hamiltonian generators proportional to different Pauli operators, the corresponding effective Gibbs states have the same eigenvalues but different modular generators. The relative entropy between these states gives a quantitative measure of the mismatch between the associated modular structures. In the minimal model this relative entropy can be computed exactly and depends only on the dimensionless parameter \(s=\beta\Delta\), where \(\Delta\) is the detector energy scale.

This also clarifies the role of information geometry. The Bogoliubov--Kubo--Mori metric arises from the second-order expansion of relative entropy, whereas the Bures metric captures operational distinguishability through fidelity \citep{Petz1996}. For commuting thermal deformations these structures coincide in the expected classical limit, but for non-commuting deformations they differ. This difference provides a quantitative distinction between entropic irreversibility and state distinguishability in accelerated detector systems.

The aim of the paper is therefore not to claim that time itself emerges from the detector model. A proper-time parameter is assumed throughout, as in standard relativistic detector theory. The result is more specific: KMS structure and non-commuting detector couplings determine when the ordering associated with that parameter becomes measurable in the detector state. This gives an operational interpretation of modular irreversibility in accelerated quantum systems.

The paper is organized as follows. Section 2 reviews the detector and KMS framework needed for the analysis. Section 3 introduces the model with non-commuting detector couplings and identifies the ordering-dependent contribution to the reduced state. Section 4 specializes to uniformly accelerated motion and shows how the KMS condition produces a thermally weighted ordering asymmetry. Section 5 relates the result to relative entropy, the Bogoliubov--Kubo--Mori metric, and modular generators of effective Gibbs states. Section 6 summarizes the conclusions and discusses extensions.

\section{Algebraic and physical framework}

We collect the elements required for the analysis, focusing on the structures that directly enter the detector dynamics.

Let $\mathcal{A}$ be the algebra of observables of a quantum field and let $\omega$ be a state on $\mathcal{A}$. For a scalar field $\phi$, the two-point correlation function is given by
\[
W(x,x') = \omega(\phi(x)\phi(x')).
\]
When evaluated along a worldline $x(\tau)$, this defines a function $W(\tau,\tau')$ that governs the response of localized detectors. In perturbation theory, detector transition probabilities and reduced states depend only on this two-point function \citep{BirrellDavies1982,Takagi1986,LoukoSatz2006}.

A state is said to satisfy the KMS condition at inverse temperature $\beta$ with respect to a flow parameter $\tau$ if correlation functions admit an analytic continuation and satisfy
\[
\omega(A(\tau)B(\tau')) = \omega(B(\tau')A(\tau + i\beta))
\]
for suitable observables $A$ and $B$ \citep{Haag1996,BratteliRobinson1997}. This condition provides a representation-independent characterization of thermal equilibrium.

A physically relevant realization arises for uniformly accelerated observers. For a trajectory with constant proper acceleration $a$, the restriction of the Minkowski vacuum to the corresponding worldline satisfies the KMS condition with inverse temperature
\[
\beta = \frac{2\pi}{a}.
\]
This is the Unruh effect \citep{Unruh1976,Davies1975}, which can be understood algebraically through the Bisognano--Wichmann theorem \citep{Bisognano1975,Bisognano1976}.

We model the detector as a two-level system interacting with the field through an Unruh--DeWitt coupling. The interaction Hamiltonian in the interaction picture is
\[
H_I(\tau) = \lambda\, \chi(\tau)\, \mu(\tau)\, \phi(x(\tau)),
\]
where $\lambda$ is a coupling constant, $\chi(\tau)$ is a switching function, and $\mu(\tau)$ acts on the detector Hilbert space. The corresponding evolution operator is
\[
U = \mathcal{T} \exp\left(-i \int d\tau\, H_I(\tau)\right),
\]
with $\mathcal{T}$ denoting time ordering. For an initial product state $\rho_D \otimes \rho_\phi$, the reduced detector state after the interaction is
\[
\rho_D' = \mathrm{Tr}_\phi \left( U (\rho_D \otimes \rho_\phi) U^\dagger \right).
\]
Expanding perturbatively, the leading contributions depend on integrals of the Wightman function along the trajectory \citep{BirrellDavies1982,Takagi1986,LoukoSatz2006}.

The framework described above provides the connection between field correlations and detector states. In the following sections, this structure is extended by allowing the detector to couple through multiple internal observables, thereby introducing non-commutativity at the level of the detector algebra.

\section{Detector model with non-commuting couplings}

We now introduce the detector setup used to probe the interplay between non-commutativity and the correlation structure of the quantum field.

We consider a two-level detector with Hilbert space $\mathcal{H}_D \simeq \mathbb{C}^2$, moving along a prescribed worldline $x(\tau)$. The detector interacts with a scalar field through couplings that act on its internal degrees of freedom. In contrast with the standard formulation, we allow the detector to couple through distinct internal observables.

Specifically, we introduce two Hermitian operators
\[
\mu_x = \sigma_x, \qquad \mu_y = \sigma_y,
\]
where $\sigma_x$ and $\sigma_y$ are Pauli matrices satisfying $[\sigma_x,\sigma_y] = 2i\sigma_z$. These operators define two independent interaction channels.

We consider interaction Hamiltonians of the form
\[
H_x(\tau) = \lambda\, \chi_x(\tau)\, \sigma_x\, \phi(x(\tau)), \qquad
H_y(\tau) = \lambda\, \chi_y(\tau)\, \sigma_y\, \phi(x(\tau)),
\]
where $\chi_x(\tau)$ and $\chi_y(\tau)$ are smooth switching functions with compact, non-overlapping support. This ensures that the detector interacts with the field through $\sigma_x$ and $\sigma_y$ in a well-defined sequence. Similar perturbative constructions with localized switching are standard in detector theory \citep{Takagi1986,LoukoSatz2006,Schlicht2004}.

Because the supports do not overlap, the time-ordered evolution operator factorizes as
\[
U_{x \to y} = U_y U_x, \qquad U_{y \to x} = U_x U_y,
\]
where
\[
U_x = \mathcal{T} \exp\left(-i \int d\tau\, H_x(\tau)\right), \qquad
U_y = \mathcal{T} \exp\left(-i \int d\tau\, H_y(\tau)\right).
\]

We compare the reduced detector states obtained from the two protocols. For an initial state $\rho_D \otimes \rho_\phi$, we define
\[
\rho_{x \to y} = \mathrm{Tr}_\phi \left( U_y U_x (\rho_D \otimes \rho_\phi) U_x^\dagger U_y^\dagger \right),
\]
\[
\rho_{y \to x} = \mathrm{Tr}_\phi \left( U_x U_y (\rho_D \otimes \rho_\phi) U_y^\dagger U_x^\dagger \right).
\]

The difference between these two states captures the dependence on the ordering of the interactions,
\[
\Delta \rho_D = \rho_{x \to y} - \rho_{y \to x}.
\]

To evaluate this quantity, we expand the evolution operators perturbatively in the coupling $\lambda$ and isolate the cross terms associated with the two interaction channels. The ordering dependence arises at second order and can be expressed in terms of the commutator of the integrated interaction operators.

A detailed derivation is provided in Appendix B. The resulting ordering-dependent contribution takes the form
\[
\Delta \rho_D
=
i\lambda^2
\int d\tau d\tau'\,
\chi_x(\tau)\chi_y(\tau')\,
G^{(1)}(\tau,\tau')\,[\sigma_z,\rho_D]
+O(\lambda^3),
\]
where
\[
G^{(1)}(\tau,\tau')
=
\omega\!\left(\{\phi(x(\tau)),\phi(x(\tau'))\}\right)
=
W(\tau,\tau')+W(\tau',\tau)
\]
is the Hadamard function along the detector trajectory.

This expression shows that the ordering asymmetry is governed by the interplay between the non-commutativity of the detector couplings and the symmetrized correlation structure of the field. The first ingredient is encoded in $[\sigma_x,\sigma_y]$, while the second is encoded in the state-dependent function $G^{(1)}$. If either ingredient is absent, the ordering-dependent contribution vanishes.

The quantity $\Delta \rho_D$ establishes that the ordering of non-commuting interactions is encoded in the reduced detector state. At this stage, the result is purely dynamical: it shows that different protocols lead to different states, but does not yet quantify their distinguishability or thermodynamic significance. This will be addressed in terms of relative entropy and information geometry in the following sections.

\section{Accelerated motion and KMS-induced asymmetry}

We now evaluate the ordering-dependent contribution to the reduced detector state for uniformly accelerated motion.

For a detector with constant proper acceleration $a$, the restriction of the Minkowski vacuum to the detector trajectory satisfies the KMS condition at inverse temperature
\[
\beta = \frac{2\pi}{a}.
\]
The corresponding Wightman function satisfies
\[
W(\tau,\tau') = W(\tau',\tau + i\beta),
\]
together with analyticity conditions \citep{Crispino2008,Haag1996,BratteliRobinson1997}.

For stationary trajectories, the correlation functions depend only on the proper-time difference. It is therefore convenient to introduce the Fourier representation of the Hadamard function,
\[
G^{(1)}(\tau,\tau')
=
\int_{-\infty}^{\infty}\frac{d\omega}{2\pi}\,
\widetilde G^{(1)}(\omega)\,e^{-i\omega(\tau-\tau')}.
\]

The KMS condition implies the detailed-balance relation for the Wightman function,
\[
\widetilde W(-\omega)=e^{-\beta\omega}\widetilde W(\omega),
\]
from which it follows that
\[
\widetilde G^{(1)}(\omega)
=
\widetilde W(\omega)+\widetilde W(-\omega)
=
\bigl(1+e^{-\beta\omega}\bigr)\widetilde W(\omega).
\]

Equivalently, introducing the spectral function
\[
\widetilde \Delta(\omega)
=
\widetilde W(\omega)-\widetilde W(-\omega),
\]
one may write
\[
\widetilde G^{(1)}(\omega)
=
\coth\!\left(\frac{\beta\omega}{2}\right)\widetilde \Delta(\omega).
\]

Substituting this into the expression for $\Delta\rho_D$, one obtains
\[
\Delta \rho_D
=
i\lambda^2[\sigma_z,\rho_D]
\int_{-\infty}^{\infty}\frac{d\omega}{2\pi}\,
\widetilde G^{(1)}(\omega)\,
\widetilde\chi_x(-\omega)\widetilde\chi_y(\omega)
+O(\lambda^3),
\]
where
\[
\widetilde\chi_j(\omega)
=
\int d\tau\,\chi_j(\tau)e^{i\omega\tau}.
\]

This expression shows that the ordering asymmetry is governed by the symmetrized correlation function of the field, with a universal thermal weighting fixed by the KMS condition. The relevant factor is not the difference between positive and negative frequencies, but their thermally weighted sum.

The KMS condition therefore plays a structural role: it fixes the relation between the two frequency sectors and induces a universal thermal scaling of the ordering asymmetry. The magnitude of the effect is controlled by the same inverse temperature $\beta$ that determines the Unruh response.

In the absence of KMS structure, no universal relation between $\widetilde W(\omega)$ and $\widetilde W(-\omega)$ exists, and the ordering dependence does not acquire a thermodynamic form. Uniform acceleration singles out a regime in which the asymmetry is governed by a well-defined local temperature.

\section{Relative entropy and information geometry}

We now quantify the ordering asymmetry in terms of relative entropy and information geometry.

For a two-level detector in local equilibrium along an accelerated trajectory, the effective states induced by distinct coupling observables can be modeled as effective Gibbs states
\[
\rho_x = \frac{e^{-\beta H_x}}{Z}, \qquad
\rho_y = \frac{e^{-\beta H_y}}{Z},
\]
with generators
\[
H_x = \Delta \sigma_x, \qquad H_y = \Delta \sigma_y,
\]
and $Z = 2\cosh(\beta \Delta)$.

These Gibbs forms should be understood as effective parametrizations capturing the local thermal response induced by the KMS structure, rather than exact stationary states of the full detector-field dynamics.

Defining $s = \beta \Delta$, these states take the form
\[
\rho_n = \frac{1}{2}(I - \tanh s \, \sigma_n).
\]

The quantum relative entropy
\[
D(\rho_y \| \rho_x) = \mathrm{Tr}(\rho_y \log \rho_y - \rho_y \log \rho_x)
\]
can be evaluated explicitly, yielding
\[
D(\rho_y \| \rho_x) = s \tanh s.
\]

This quantity depends only on the dimensionless parameter $s$ and provides a measure of the mismatch between the two effective Gibbs structures associated with different generators.

For nearby states, introduce a one-parameter family of states $\rho_\theta$ interpolating between the two effective Gibbs states, with $\theta$ parametrizing the rotation in operator space between $\sigma_x$ and $\sigma_y$. The relative entropy admits the quadratic expansion
\[
D(\rho_{\theta} \| \rho_0)
\simeq \frac{1}{2} g^{\mathrm{BKM}}_{\theta\theta} \theta^2,
\]

which for the present model yields
\[
g^{\mathrm{BKM}} = s \tanh s.
\]

By contrast, the Bures metric, derived from quantum fidelity, gives
\[
g^{\mathrm{Bures}} = \tanh^2 s.
\]

The ratio
\[
\frac{g^{\mathrm{BKM}}}{g^{\mathrm{Bures}}} = \frac{s}{\tanh s}
\]
quantifies the separation between entropic irreversibility and operational distinguishability.

At high temperature ($s \ll 1$), the two metrics coincide, while at low temperature ($s \gg 1$), they diverge. This behavior reflects the increasing role of non-commutativity in the structure of thermal states.

The relative entropy therefore provides a quantitative measure of the information retained about the ordering of interactions.
Since $D(\rho_y \| \rho_x) > 0$, the two states are distinguishable in principle. 
By standard results in quantum hypothesis testing, there exists an observable 
for which the expectation values in $\rho_x$ and $\rho_y$ differ. 
The ordering asymmetry is therefore not only a formal property of the reduced state, 
but is operationally accessible through suitable measurements on the detector.
When $D(\rho_y \| \rho_x) \neq 0$, the detector state encodes a distinction between different coupling sequences.

Combined with the dynamical result of Section 4, this shows that KMS thermality and non-commutativity jointly determine when ordering information becomes physically accessible. The parameter $s = \beta \Delta$ controls both the strength of the Unruh response and the magnitude of the ordering asymmetry.

The connection between modular structure and the thermodynamic arrow of time is most naturally expressed in terms of relative entropy. While the von Neumann entropy is conserved under unitary evolution, relative entropy with respect to a reference state provides a measure of irreversible deviation from equilibrium.

In the present setting, the accelerated trajectory selects a local KMS state $\rho_{\mathrm{KMS}}$ and hence a modular generator $K=-\log\rho_{\mathrm{KMS}}$. The associated modular flow defines a natural notion of thermal time. The physically relevant orientation of this flow is singled out by the requirement that relative entropy production be non-negative.

For the minimal non-commuting two-level model, this structure is reflected in the positivity of
\[
D(\rho_y\|\rho_x)=s\tanh s \geq 0,
\]
which measures the mismatch between the effective Gibbs states associated with different coupling observables. The arrow of thermal time can therefore be interpreted as not imposed by the external proper-time parameter alone, but selected by the KMS state through the monotonicity of relative entropy.

This provides a precise link between modular flow, thermality, and irreversibility: temporal ordering becomes physically oriented when it corresponds to increasing modular distinguishability.

\section{Discussion}

We have identified a minimal mechanism through which temporal ordering becomes operationally meaningful in relativistic quantum systems. The key ingredients are the KMS structure of the underlying state and the non-commutativity of the observables through which the system probes the field.

The mechanism can be summarized as follows. The KMS condition imposes a specific analytic structure on correlation functions, which translates into a frequency-dependent weighting of forward and backward processes. When the detector couples through non-commuting observables, this structure is reflected in the reduced detector state as a difference between distinct interaction protocols. The resulting asymmetry appears at the level of the density matrix and does not require coarse-graining or macroscopic limits.

In the minimal two-level model, the detector states associated with different internal observables are described by effective Gibbs distributions with identical spectra and different generators. The relative entropy between these states provides a natural measure of their difference. Its explicit form shows that the same parameter controlling the Unruh response also governs the strength of the ordering asymmetry.

The comparison between the Bogoliubov--Kubo--Mori and Bures metrics clarifies the distinction between entropic and operational aspects of this effect. In commuting directions, these metrics coincide, reflecting the classical character of thermal fluctuations. In non-commuting directions, they differ, and this difference is controlled by the same parameter that determines the thermal response. The resulting structure links distinguishability, irreversibility, and local thermal scale in a unified way.

The analysis presented here is based on perturbation theory and localized detector models. It therefore inherits the assumptions of that framework, including the use of switching functions and the restriction to leading nontrivial order in the coupling. Extending the results beyond this regime would require a non-perturbative treatment of detector dynamics or an explicit construction within algebraic quantum field theory.

The framework suggests several directions for further investigation. One natural extension is to consider more general detector systems, including higher-dimensional internal Hilbert spaces or continuous-variable models, where the structure of non-commuting observables is richer. Another direction is to study different spacetime settings in which KMS states arise, such as stationary spacetimes with horizons, where similar mechanisms may be present.

The relation between modular structure and information geometry may also be explored further. In particular, the role of relative entropy as a measure of modular mismatch suggests a possible link with other contexts in which modular Hamiltonians play a central role, such as quantum field theory in curved spacetime and quantum information theory.

The results show that KMS thermality, modular structure, and non-commutativity jointly determine when temporal ordering becomes observable and when it acquires a preferred direction. In this sense, the arrow of thermal time emerges as the direction along which relative entropy with respect to the KMS state increases.

A natural extension of this work is to analyze whether similar ordering asymmetries arise in other settings where KMS states are present, such as stationary spacetimes with horizons. It would also be of interest to explore non-perturbative formulations of the effect and its relation to modular Hamiltonians in algebraic quantum field theory.

\appendix

\section{Derivation of relative entropy for non-commuting Gibbs states}

In this appendix we derive the closed-form expression for the relative entropy between the two effective Gibbs states introduced in Section 5.

We consider a two-level system with density matrices
\[
\rho_x = \frac{e^{-s \sigma_x}}{2\cosh s}, \qquad
\rho_y = \frac{e^{-s \sigma_y}}{2\cosh s},
\]
where $s = \beta \Delta$.

Using the identity for Pauli matrices
\[
e^{-s \sigma_n} = \cosh s \, I - \sinh s \, \sigma_n,
\]
one obtains
\[
\rho_n = \frac{1}{2}\left(I - \tanh s \, \sigma_n\right).
\]

The logarithm of the density matrix is
\[
\log \rho_n = -\log(2\cosh s)\, I - s \sigma_n.
\]

The relative entropy is defined as
\[
D(\rho_y \| \rho_x) = \mathrm{Tr}(\rho_y \log \rho_y) - \mathrm{Tr}(\rho_y \log \rho_x).
\]

Substituting the expressions above gives
\[
D(\rho_y \| \rho_x)
= \mathrm{Tr}\left[\rho_y \left(-s\sigma_y + s\sigma_x\right)\right].
\]

Using
\[
\mathrm{Tr}(\rho_y \sigma_y) = -\tanh s, \qquad
\mathrm{Tr}(\rho_y \sigma_x) = 0,
\]
we obtain
\[
D(\rho_y \| \rho_x) = s \tanh s.
\]

\section{Perturbative derivation of the ordering asymmetry}

We derive the second-order ordering-dependent contribution to the reduced detector state for the two sequential interaction protocols introduced in Section 3.

Let
\[
A_x = \int d\tau\, H_x(\tau), \qquad
A_y = \int d\tau\, H_y(\tau),
\]
and
\[
B_x = \int d\tau d\tau'\, \mathcal{T}\big(H_x(\tau)H_x(\tau')\big), \qquad
B_y = \int d\tau d\tau'\, \mathcal{T}\big(H_y(\tau)H_y(\tau')\big).
\]

To second order in the coupling, the individual evolution operators are
\[
U_x \approx 1 - iA_x - \frac{1}{2}B_x, \qquad
U_y \approx 1 - iA_y - \frac{1}{2}B_y.
\]

Since the supports of the switching functions do not overlap, the two ordered protocols are
\[
U_{x\to y} \approx 1 - i(A_x+A_y) - \frac{1}{2}(B_x+B_y) - A_yA_x,
\]
\[
U_{y\to x} \approx 1 - i(A_x+A_y) - \frac{1}{2}(B_x+B_y) - A_xA_y.
\]

For an initial product state $\rho_D\otimes\rho_\phi$, the reduced detector states are
\[
\rho_{x\to y}=\mathrm{Tr}_\phi\!\left(U_{x\to y}(\rho_D\otimes\rho_\phi)U_{x\to y}^\dagger\right),
\]
\[
\rho_{y\to x}=\mathrm{Tr}_\phi\!\left(U_{y\to x}(\rho_D\otimes\rho_\phi)U_{y\to x}^\dagger\right).
\]

At order $\lambda^2$, the first-order contributions and the quadratic sandwich term are identical for both protocols. The difference therefore arises entirely from the ordered cross terms and can be written as
\[
\Delta\rho_D = \rho_{x\to y} - \rho_{y\to x}
= \mathrm{Tr}_\phi\!\left(\bigl[[A_x, A_y],\, \rho_D \otimes \rho_\phi\bigr]\right)
+ O(\lambda^3).
\]

Using
\[
H_x(\tau)=\lambda\,\chi_x(\tau)\,\sigma_x\,\phi(x(\tau)),
\qquad
H_y(\tau')=\lambda\,\chi_y(\tau')\,\sigma_y\,\phi(x(\tau')),
\]
we compute the commutator
\[
[A_x,A_y]
=
\lambda^2 \int d\tau d\tau'\,\chi_x(\tau)\chi_y(\tau')
\,[\sigma_x\phi(x(\tau)),\,\sigma_y\phi(x(\tau'))].
\]

Using
\[
[A\otimes B,\,C\otimes D]=[A,C]\otimes BD + CA\otimes[B,D],
\]
together with
\[
[\sigma_x,\sigma_y]=2i\sigma_z,
\qquad
\sigma_y\sigma_x=-i\sigma_z,
\]
one finds
\[
[\sigma_x\phi(x(\tau)),\,\sigma_y\phi(x(\tau'))]
=
i\sigma_z\,\{\phi(x(\tau)),\phi(x(\tau'))\}.
\]

Therefore,
\[
[A_x,A_y]
=
i\lambda^2
\int d\tau d\tau'\,\chi_x(\tau)\chi_y(\tau')\,
\sigma_z\,\{\phi(x(\tau)),\phi(x(\tau'))\}.
\]

Tracing over the field degrees of freedom gives
\[
\Delta \rho_D
=
i\lambda^2
\int d\tau d\tau'\,\chi_x(\tau)\chi_y(\tau')\,
G^{(1)}(\tau,\tau')\,[\sigma_z,\rho_D]
+ O(\lambda^3),
\]
where
\[
G^{(1)}(\tau,\tau')
=
\omega\!\left(\{\phi(x(\tau)),\phi(x(\tau'))\}\right)
=
W(\tau,\tau')+W(\tau',\tau)
\]
is the Hadamard function.

For stationary trajectories, one may introduce the Fourier representation
\[
G^{(1)}(\tau,\tau')
=
\int_{-\infty}^{\infty}\frac{d\omega}{2\pi}\,
\widetilde G^{(1)}(\omega)e^{-i\omega(\tau-\tau')}.
\]

If the state satisfies the KMS condition,
\[
\widetilde W(-\omega)=e^{-\beta\omega}\widetilde W(\omega),
\]
then
\[
\widetilde G^{(1)}(\omega)
=
\widetilde W(\omega)+\widetilde W(-\omega)
=
\bigl(1+e^{-\beta\omega}\bigr)\widetilde W(\omega)
=
\coth\!\left(\frac{\beta\omega}{2}\right)\widetilde\Delta(\omega),
\]
where $\widetilde\Delta(\omega)=\widetilde W(\omega)-\widetilde W(-\omega)$.

This shows explicitly that the ordering asymmetry is governed by the symmetrized correlation function of the field, with a thermal weighting fixed by the KMS condition.

\end{document}